\documentclass[prl,twocolumn,aps,twocolumn]{revtex4-1}

\usepackage{amsmath,amssymb,graphicx}
\usepackage{epsf}
\usepackage{color}
\usepackage{latexsym} 
\usepackage{amsmath,amsthm}
\usepackage{ifpdf}
\usepackage{epstopdf}
\usepackage{dcolumn}% Align table columns on decimal point
\usepackage{bm}% bold math
\usepackage{braket}
\usepackage{natbib}

\begin{document}

\def \beq{\begin{equation}}
\def \eeq{\end{equation}}
\def \bea{\begin{eqnarray}}
\def \eea{\end{eqnarray}}
\def \bem{\begin{displaymath}}
\def \eem{\end{displaymath}}
\def \P{\Psi}
\def \Pd{|\Psi(\boldsymbol{r})|}
\def \Pds{|\Psi^{\ast}(\boldsymbol{r})|}
\def \Po{\overline{\Psi}}
\def \bs{\boldsymbol}
\def \bl{\bar{\boldsymbol{l}}}

\title{Squeezing in a nonlocal photon fluid}
\vspace{0.5cm}

\author{M. C. Braidotti$^{1,2\star}$, A. Mecozzi$^1$ and C. Conti$^{2,3}$}
\affiliation{\small
$^1$ Department of Physical and Chemical Sciences, University of L'Aquila, Via Vetoio 10, I-67010 L'Aquila, Italy\\
$^2$ Institute for Complex Systems, National Research Council (ISC-CNR), Via dei Taurini 19, 00185 Rome, Italy\\
$^3$ Department of Physics, University Sapienza, Piazzale Aldo Moro 5, 00185 Rome, Italy\\
$^*$Corresponding author: mariachiara.braidotti@isc.cnr.it}
\date{\today}

\begin{abstract}
Quantum fluids of light are an emerging tool employed in quantum many-body physics. Their amazing properties and versatility allow using them in a wide variety of fields including gravitation, quantum information and simulation. However the implications of the quantum nature of light in the nonlinear optical propagation are still missing many features. We theoretically predict classical spontaneous squeezing of a photon fluid in a nonlocal nonlinear medium. 
By using the so called Gamow vectors, we show that the quadratures of a coherent state get squeezed and that a maximal squeezing power exists. 
Our analysis holds true for temporal and spatial optical propagation in highly nonlocal regime. These results open a new scenario in quantum photon fluids and may lead to novel applications in fields like metrology and analogues of quantum gravity. 
\end{abstract}


\maketitle

%%%%%%%%%%%%%%%%%%%%%%%%%%%%%%%%%%%%%%%%%%%%%%%%%%%%%%%%%%%

A recent forefront topic of research is the study of quantum fluids, which arise in a wide variety of fields ranging from condensed matter to particle physics (see \cite{Carusotto2013,Vocke15,Carusotto2014,Carusotto2015} and references therein). Quantum fluids are many particle systems in which the average particle distance becomes comparable or smaller than the thermal de Broglie wavelength. In these cases the statistical properties of the system becomes fundamental in describing the properties of the fluid. One of the most popular quantum fluid is the Bose-Einstein condensate where a great number of particles share the same energy state \cite{Weitz2010,Weitz2012}.
Lately emerged the possibility of studying propagating classical light as a quantum fluid of photons where the photon-photon interactions are mediated by a nonlinear optical medium. A particularly intriguing scenario is the study of nonlocal mechanism both in the temporal and spatial domain \cite{Calvanese2014,ContiBiancalana2010,Ghofraniha07,Mecozzi:96}. Nonlocality allows interesting phenomena such as analogue boson stars described by the Newton-Schr\"odinger equation\cite{Liebling,Faccio2016natcom}, large scale coherence and condensation processes \cite{Sun2012,Picozzi06,Weitz2012,WeitzNAT2010,WeitzNatPhys2010,connaughton:263901}. In this scenario, there are numerous unexplored directions and missing features mostly in linking quantum mechanics and fluids of light. These features might be relevant for application in quantum information and simulation. \\
In this Letter, we predict the presence of squeezing in nonlocal photon fluids.
Squeezed states are pure quantum states, which proved to play an important role in modern quantum optics \cite{Walls1983,WallsBook}. 
These quantum states were first discovered in 1927 by Kennard\cite{Kennard27}, however their mathematical properties were investigated in 70s and 80s \cite{Stoler1970,Stoler1971,Fisher1984,Ma1990}.
The use of squeezed states in cryptography, quantum computation and gravitational wave detection has attracted great attention \cite{Gisin2002,Ladd2010,Ligo2011,Chua2014,Shnabel2010}. Nowadays the challenge is reaching maximal squeezing that can be implemented in gravitational waves detectors \cite{Vahlbruch2016}.\\
Here, we show that the squeezing operator $\hat S(\zeta)$ naturally occurs in the evolution of a photon fluid with nonlocal and/or non-instantaneous nonlinearity. We hence theoretically predict that photon fluids are squeezed states. 
A leading ingredient in our theory is the link with the reversed harmonic oscillator and the eigenstates of $\hat S(\zeta)$  given by the so called Gamow vectors (Gvs) \cite{gamow1928ZurQuadesAto,Chruscinski2004squeeze,Prigogine77,nucl-th/9902076v1,Ford1959} that we illustrate in the following\cite{Gentilini2015glauber,Gentilini2015}.

%%%%%%%%%%%%%%%%%%%%%%%%%%%%%%%%%%%%%%%%%%%%%%%%%%%%%%%%%%%
We start analyzing the spatially nonlocal non\-li\-near optical propagation in the one-dimensional paraxial case. Below we take into account also a pulse evolution in a non-instantaneous medium. We consider a linearly polarized Gaussian beam with wavelength $\lambda$ and amplitude $A$ in a defocusing nonlinear medium along the $Z$ direction:
\begin{equation}
2ik\partial_Z A+\partial_X^2 A +2k^2\frac{\Delta n[|A|^2](X)}{n_0}A=0,
\label{nls}
\end{equation}
where $n_0$ is the linear refractive index of the medium and $k=2\pi/\lambda$ is the wavenumber. $A$ is normalized such that $|A|^2=I$ is the intensity. In Eq. (\ref{nls}) the perturbation to the refractive index $\Delta n[I](X)$ can be written as
\begin{equation}
\Delta n[I](X)=n_2\int{G_2(X-X')I(X')dX'},
\label{dn}
\end{equation}
where $n_2$ is the nonlinear coefficient and $G_2$ is the kernel function for an exponential nonlocality $G_2(X)=\mbox{exp}(-|X|/L_{nloc})/2L_{nloc}$.
We write Eq.(\ref{nls}) in terms of the adimensional variables $z=Z/Z_d$ with $Z_d=kW_0^2$ and $x=X/W_0$, being $W_0$ the Gaussian beam waist:
\begin{equation}
i\partial_z \psi+\frac{1}{2}\partial_x^2 \psi -PK(x)\ast|\psi|^2\psi=0,
\label{nls_ad}
\end{equation}
where $\psi=A\sqrt{P_{MKS}}/W_0$ and $P=P_{MKS}/P_{ref}$ with $P_{ref}=\lambda^2/4\pi^2n_0|n_2|$. $K(x)=W_0G_2(xW_0)=exp(-|x|/\sigma)/(2\sigma)$ is the nonlocal function with $\sigma=L_{nloc}/W_0$ the degree of nonlocality.
In the highly nonlocal approximation, i.e., when the nonlocality length is much wider than the beam waist $(L_{nloc}>W_0)$ or $\sigma>1$, one can write the convolution as $\kappa(x)\simeq K(x)\ast|\psi|^2$. Eq.(\ref{nls_ad}) becomes $i\partial_z \psi=\hat H \psi$, with $\hat H=\frac{1}{2}\hat p^2 + V(x)$, $\hat p=-i\partial_x$ and $V(x)=P\kappa(x)$. A series expansion of $V(x)$ at $x=0$ gives $\kappa(x)\simeq\kappa_0^2-\frac{1}{2}\kappa_2^2 x^2$ with $\kappa_0=1/\sqrt{2\sigma}$ and $\kappa_2=(\sigma\sqrt[4]{\pi})^{-1}$. We have $\hat H = P\kappa_0^2 + \hat H_{RHO}$, where
\begin{equation}
\hat H_{RHO} = \frac{\hat p^2}{2} - \frac{\gamma^2 \hat x^2}{2}
\label{Hrho}
\end{equation} 
is the Hamiltonian of a reversed harmonic oscillator, with $\gamma^2=P\kappa_2^2=P/\sigma^2\sqrt{\pi}$.
Equation (\ref{Hrho}) is typically studied in scaled coordinates with $\gamma=1$ in which the squeezed quadratures are immediately identified. A key difference in our case is the fact that the $\gamma$ parameter depends on the beam power $P$ because of the nonlinear dynamics. As a result the squeezing quadratures, defined below, rotates when varying the input flux. In the following, we will report a theory for such a nonlinear squeezing.\\ 
First, we let $\psi=\mbox{exp}(-iP\kappa_0^2z)\Psi$ and obtain $ i\partial_z \Psi=\hat H_{RHO} \Psi$.
This shows that a nonlocal photon fluid is actually described by the reversed harmonic oscillator Hamiltonian. One can realize that the evolution is hence given by the squeezing operator as follows. 
The evolved wave function $\Psi(z)$ can be written in terms of the propagator of quantum mechanics $\hat U(z)=\mbox{exp}(-i\hat H_{RHO} z)$ as $\Psi(z)=\hat U(z)\Psi(0)$. The key-point is that $\hat U(z)$ can be expressed as the squeezing operator $\hat S(\zeta)$ with $\zeta=r e^{i\theta}$, where $r$ is the \emph{squeezing parameter}, as follows:
the operator $\hat S(\zeta)$ is  
\begin{equation}
\hat S(\zeta)=\mbox{exp}\left[\frac{1}{2}\left(\zeta^* \hat a^2-\zeta \hat a^{\dag^2} \right)\right],
\label{squeeze}
\end{equation}
and we consider   
\begin{equation}
\hat H_{\zeta}=\frac{i}{2z}\left(\zeta^* \hat a^2 - \zeta \hat a^{\dag^2} \right).
\label{squeeze}
\end{equation}
The operator $\hat H_{RHO}=\hat R^{\dag}(\varphi)\hat H_{\zeta} \hat R(\varphi)$, with $\hat R(\varphi)$ a single mode rotation, is unitedly equivalent to $\hat H_{\zeta}$ \cite{Chruscinski2004squeeze}.
Introducing the creation and annihilation operators:
\begin{equation}
\hat a=\frac{\hat u+i \hat v}{\sqrt{2}},\qquad \hat a^{\dag}=\frac{\hat u-i \hat v}{\sqrt{2}},
\end{equation}
we find that 
\begin{equation}
\hat H_{\zeta}=-\frac{\gamma}{2}(\hat u\hat v+\hat v\hat u)
\label{Hr}
\end{equation}
which is the RHO Hamiltonian in Eq. (\ref{Hrho}) by the fol\-lo\-wing canonical transformations
\begin{equation}
\hat u=\frac{\gamma \hat x-\hat p}{\sqrt{2\gamma}},\qquad \hat v=\frac{\gamma \hat x+\hat p}{\sqrt{2\gamma}}.
\end{equation}
The quadratures $\hat u$ and $\hat v$ are squeezed during evolution, i.e., one decreases exponentially below the coherent Gaussian limit, while the other increases: the squeezing parameter is $r=\gamma z$ and the angle is $\theta=0$.\\
To quantify the squeezing occurring during the evolution of a Gaussian wave-packet we adopt Gvs, which are the generalized eigensolutions of the RHO:
$\hat H_{RHO}\mathfrak{f}_n^{\pm}=\pm E_n\mathfrak{f}_n^{\pm}$,
with imaginary eigenvalues $E_n=i\gamma(n+1/2)$. Gvs are also the eigenstates of the squee\-zing operator $\hat S(r)$ with real eigenvalues $s_n^{\pm}=\mbox{exp}\left[\pm r \left(n+\frac{1}{2}\right)\right]$ \cite{Chruscinski2004squeeze}
\begin{equation}
\hat S(r)\mathfrak{f}_n^{\pm}=e^{\mp i E_n z}\mathfrak{f}_n^{\pm}=s_n^{\pm}\mathfrak{f}_n^{\pm}.
\end{equation}
On has $ \mathfrak{f}_n^+=\frac{u^n}{\sqrt{n!}}$.
Our initial condition is a Gaussian beam $\psi_0(x)=(\pi)^{-1/4}\mbox{exp}\left[-x^2/2\right]$ in the $x$ space. $\psi_0(x)$ in the $u$ space reads \cite{Marcucci16}
\begin{equation}
\psi_{0}(u)=(-1)^{1/8}\sqrt[4]{\frac{2\gamma}{\pi}}\frac{e^{\frac{-u^2(1-i\gamma)}{2(\gamma-i)}}}{\sqrt{1+i\gamma}}.
\label{psiu_new}
\end{equation}
Equation (\ref{psiu_new}) can be expanded in series of Gvs $\mathfrak{f}^{+}_{n}$:
\begin{equation}
\psi_{0}(u)=c \sum_{n=0}^{\infty}\left[\frac{(-1)(1-i\gamma)}{(\gamma-i)} \right]^n\sqrt{\frac{(2n-1)!!}{(2n)!!}}\mathfrak{f}^+_{2n}(u),
\end{equation}
with $c =(-1)^{1/8}\sqrt[4]{\frac{2\gamma}{\pi(1+i\gamma)^2}}$. 
The squeezing operator $S(r)$, hence, acts as 
\begin{equation}
\begin{aligned}
S(r)\psi_0(u)&=c e^{r/2}\sum_{n=0}^{\infty}\left[\frac{(-1)(1-i\gamma)}{(\gamma-i)} \right]^n\times\\
&\times\sqrt{\frac{(2n-1)!!}{(2n)!!}}e^{2 r n}\mathfrak{f}^+_{2n}(u)=e^{r/2} \psi_0(ue^{r}).
\label{evolved}
\end{aligned}
\end{equation}
%%%%%%%%%%%%%%%%%%%%%%%%%%%%%%% FIGURE 1 %%%%%%%%%%%%%%%%%%%%%%%%%%%%%%%%%%%%%
\begin{figure}[t!]
	\centering
		\includegraphics[scale=0.5]{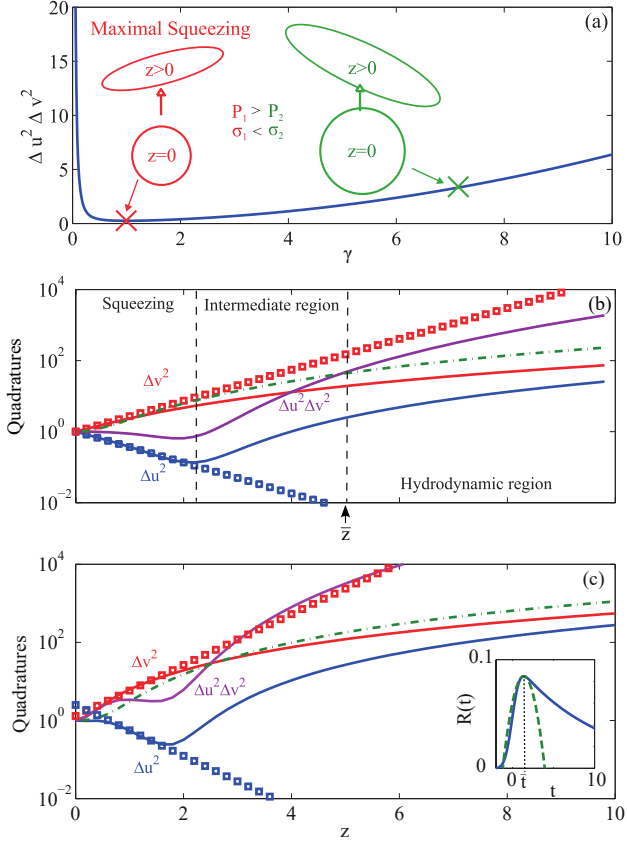}
	\centering
	\caption{(Color online) (a) Uncertainty principle $\Delta u^2 \Delta v^2$ behavior as function of $\gamma$. The inset shows that the phase space distribution area changes with $\gamma$ and hence with $P$ and $\sigma$, while the phase space distribution form varies from circular to elliptical during the propagation. (b) Normalized quadratures $u$ (blue line) and $v$ (red line) uncertainties and uncertainty principle $\Delta u^2\Delta v^2$ (magenta line). Squares represent the theoretical behavior for the two normalized quadratures $u$ (blue squares) and $v$ (red squares) in semi-log-scale after Eq.(\ref{uncertainty}) with $P=100$ and $\sigma=15$. Green dot-dashed curve provides the expected result in the hydrodynamical regime after Eq. (\ref{hydrodyn}). Dashed line evidence the $3$ region limits. (c) Same as (b) for the temporal case with $P=100$ and $T=10$. The inset is the convolution integral $\mathcal{R}(t)$ behavior as function of time (blue line). The green dashed line is the series expansion trend around the convolution maximum $\bar t$.}
\label{fig:1}
\end{figure}
%%%%%%%%%%%%%%%%%%%%%%%%%%%%%%%%%%%%%%%%%%%%%%%%%%%%%%%%%%%%%%%%%%%%%%%%%%%%%%%%%%%%%%%%
From Eq. (\ref{evolved}), we calculate the quadrature uncertainties $\Delta u^2$ and $\Delta v^2$ for the evolved state (\ref{evolved})
\begin{equation}
\begin{aligned}
\Delta u^2&=\Braket{S(r)\psi_0(u)|\hat u^2|S(r)\psi_0(u)}=\frac{e^{-2r}}{4}\left(\frac{1+\gamma^2}{\gamma}\right),\\
\Delta v^2&=\frac{e^{2r}}{4}\left(\frac{1+\gamma^2}{\gamma}\right),
\label{uncertainty}
\end{aligned}
\end{equation}
The $\hat u$ uncertainty decreases at the cost of the correspon\-ding increase in $\Delta v$. 
The uncertainty principle reads as 
\begin{equation}
\Delta u^2\Delta v^2=\frac{(1+\gamma^2)^2}{16\gamma^2}.
\label{du*dv}
\end{equation}
Remarkably, this generalized uncertainty principle $\Delta u^2 \Delta v^2$  predicts both that the squeezing degree depends on $\gamma$ and hence on the beam power and that, at fixed initial waist, $\Delta u\Delta v$ changes with the degree of nonlocality and the beam power. 
Figure \ref{fig:1}(a) reports $\Delta u\Delta v$ as function of $\gamma$, showing that a maximal degree of squeezing exists and corresponds to $\gamma=1$. For $\Delta x$ and $\Delta p$  one has
\begin{equation}
\begin{aligned}
&\Delta x^2=\frac{\gamma^2-1}{4\gamma^2}+\frac{1+\gamma^2}{4\gamma^2}\cosh(2r),\\
&\Delta p^2=\frac{1-\gamma^2}{4}+\frac{1+\gamma^2}{4}\cosh(2r),\\
&\Delta x^2\Delta p^2=\frac{(1+\gamma^2)^2}{16\gamma^2}\cosh^2(2r)-\frac{(\gamma^2-1)^2}{16\gamma^2}.
\label{uncertainty2}
\end{aligned}
\end{equation}
We remark that the squeezing in Eqs. (\ref{uncertainty}), (\ref{du*dv}) and (\ref{uncertainty2}) is obtained by the exact solution of the highly nonlocal nonlinear Schr\"odinger equation.\\
Notably, we now show that no squeezing is predicted in the hydrodynamical approximation commonly adopted in solving the nonlinear Schr\"oedinger equation (see, e.g.,\cite{ContiBiancalana2010}). In order to compute the hydrodynamical limit of the two quadratures $\hat u$ and $\hat v$, we define $\varepsilon=\sqrt{L_{nl}/Z_d}$, where $L_{nl}$ and $Z_d$ are the adimensional nonlinear and diffractive lengths respectively. $\varepsilon$ is a small parameter which accounts the competition between diffraction and nonlinearity. In the adimensional coordinates $\chi=x\varepsilon$ and $\rho=z\varepsilon$,  Eq. (\ref{nls_ad}) reads as
\begin{equation}
i\varepsilon\partial_{\rho} \psi+\frac{1}{2}\varepsilon^2\partial_{\chi}^2 \psi -PK(\chi/\varepsilon)\ast|\psi|^2\psi=0
\end{equation}
and the quadratures $\hat u$ and $\hat v$ become
\begin{equation}
\begin{aligned}
\hat u=\frac{\gamma x-\hat p}{\sqrt{2\gamma}}\quad\longrightarrow 
\frac{\frac{\gamma\chi}{\varepsilon}+i\varepsilon\partial_{\chi}}{\sqrt{2\gamma}}\\
\hat v=\frac{\gamma x+\hat p}{\sqrt{2\gamma}}\quad\longrightarrow 
\frac{\frac{\gamma\chi}{\varepsilon}-i\varepsilon\partial_{\chi}}{\sqrt{2\gamma}}.
\end{aligned}
\end{equation}
For $\varepsilon \ll 1$, with $\psi=A e^{i\phi/\varepsilon}$, the Wentzel–Kramers–Brillouin (WKB) approach is applicable and we obtain, as $\varepsilon\rightarrow 0$, the uncertainties of the two quadratures
\begin{equation}
\begin{aligned}
\Delta u^2\rightarrow\frac{\gamma}{2}\Delta x^2,\\
\Delta v^2\rightarrow\frac{\gamma}{2}\Delta x^2.
\label{hydrodyn}
\end{aligned}
\end{equation}
This result implies that in the hydrodynamical regime the quadratures $\hat u$ and $\hat v$ have the same evolution.\\
The formalism just developed shed light on the presence of three different regimes in the nonlinear nonlocal photon fluids propagation. In the first part of the propagation squeezing takes place. Then, there is an intermediate region in which the squeezing stops and both the quadratures start increasing. In the third region the propagation becomes highly nonlinear.  \\    
The presence of an intermediate region, strictly connected with the highly nonlocal approximation (HNA), can be also proved analytically. When the HNA does not hold true anymore, the nonlinearity starts having a dominant role in the wave propagation, which enters the nonlinear regime. 
This happens when the beam waist $W(z)=\Delta  x$ becomes comparable to the degree of nonlocality $\sigma$:
$%\begin{equation} 
\Delta x^2 = \frac{\gamma^2-1}{4\gamma^2}+\frac{1+\gamma^2}{4\gamma^2}\cosh(2r) \simeq \sigma^2.
$%\end{equation}
The value of $z$ at which the transition between the squeezing and the highly nonlinear regimes happens is around 
\begin{equation} 
 \bar z = \frac{\mbox{log}\left[\frac{\gamma^2(4\sigma^2-1)+1}{1+\gamma^2} + \sqrt{\left(\frac{\gamma^2(4\sigma^2-1)+1}{1+\gamma^2}\right)^2 - 1} \right]}{2\gamma}.
\label{zbarra}
\end{equation}
Equation (\ref{zbarra}) is successfully compared with numerical simulation below. 
%%%%%%%%%%%%%%%%%%%%%%%%%%%%%%% FIGURE 2 %%%%%%%%%%%%%%%%%%%%%%%%%%%%%%%%%%%%%
\begin{figure}[t!]
	\centering
		\includegraphics[scale=0.5]{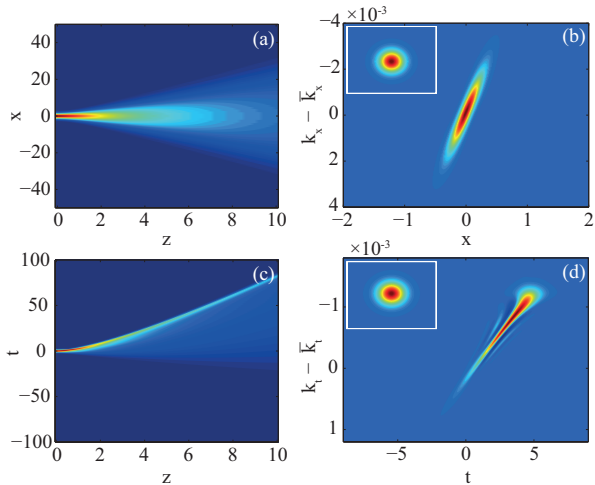}
	\centering
		\caption{(a) Numerical solution of Eq.(\ref{nls_ad}) with $P = 100$ and $\sigma = 15$; (b) Wigner function calculated after Eq.(\ref{nls_ad}) of the evolved beam in panel (a) at $z=2$, with $\bar k_x=257$; the inset shows the Wigner function of the initial condition of panel (a).  (c) Numerical solution of Eq.(\ref{tnls}) with $P = 100$ and $T = 10$; (d) the same as (b) for the evolved beam in panel (c), with $\bar k_t=128$.}
	\label{fig:2}
\end{figure}
%%%%%%%%%%%%%%%%%%%%%%%%%%%%%%%%%%%%%%%%%%%%%%%%%%%%%%%%%%%%%%%%%%%%%%%%%%%%%%%%%%%%%%%%

In order to test our theory, we simulate a Gaussian beam $\psi(x,0)=\psi_0(x)$ propagating according to Eq.(\ref{nls_ad}) in a nonlocal nonlinear medium. Figure \ref{fig:1}(b) shows the normalized uncertainty of the quadratures $\hat u$ and $\hat v$ trend. We observe the presence of the three predicted regions. In the initial stage of propagation, the quadrature $\hat u$ is squeezed, i.e., its uncertainty value decreases with respect to $1$, which is the normalized uncertainty for the coherent state, while $\Delta v$ 
diverges exponentially. Fig. \ref{fig:1}(b) shows that the RHO-approximation of Eq. (\ref{nls_ad}) holds true, and $\Delta u$ and $\Delta v$ show exponential trends. During evolution, the system reaches a maximum squeezing. After that, the two quadratures follow the same dynamics. Note that the product $\Delta u \Delta v$ stays constant in the squeezing region, as expected form Eq. (\ref{du*dv}). These numerical results show that the hydrodynamical ap\-pro\-xi\-ma\-tion fails to catch the squeezing dynamics. Only after the maximal squeezing has been reached, the two quadratures tend to have the same trend as predicted by Eq. (\ref{hydrodyn}): for large propagation distances the two quadratures tend to the same limit given by the green dot-dashed line in Fig. \ref{fig:1}(b). The transition point $\bar z$, calculated after Eq. (\ref{zbarra}), corresponds with the limit of the intermediate region where the highly nonlocal approximation stops holding true. The Gamow approach is more accurate for what concerns the two quadratures; the hydrodynamical approach fails at the lowest order because it neglects the derivative of the density $|\psi|^2$ and only accounts for the dynamics of the phase. %Our theory hence surpasses previous theoretical approaches. NELLA RISPOSTA

The scenario also holds true for temporal photon fluids. We consider the defocusing non-instantaneous nonlinear Shr\"{o}dinger equation in the dimensionless form 
\begin{equation}
i\partial_z\psi+\partial_t^2 \psi -\psi\int_{-\infty}^{+\infty}R(t-t')|\psi(t')|^2dt'=0,
\label{tnls}
\end{equation}
where $R(t)=(1/T)\exp(-t/T)\Theta(t)$ is the medium response function and $T$ is the medium response time. 
$\Theta(t)$ is the Heaviside function which guarantees time causa\-lity. The response function $R(t)$ is normalized such that $\int{R(t)dt}=1$ \cite{Conti2010_linearons,Skryabin03}.
In the highly non-instantaneous limit, i.e. medium response time longer than the pulse duration $(T\gg t_0)$, Eq. (\ref{tnls}) becomes effectively linear:
\begin{equation}
i\partial_z\psi+\partial_t^2 \psi -\mathcal{E}R(t)\psi=0,
\label{linear-tnls}
\end{equation}
where $\mathcal{E}=\int{|\psi(t)|^2dt}$ is the pulse energy. 
Equation (\ref{tnls}) holds true in fibers with focusing nonlinearity in the normal dispersion regime. The highly non-instantaneous regime describes liquid filled hollow core fibers as those studied in \cite{Conti2010_linearons}. In analogy with the spatial case, the temporal dynamics of a nonlocal photon fluid can be described both by the hydrodynamical \cite{ContiBiancalana2010} and the Gamow vectors approach. In particular, Eq. (\ref{tnls}) admits  eigenfunctions of the form $\psi=\phi_E e^{iEz}$, as in \cite{Conti2010_linearons}. These solutions lead to Gamow vectors with imaginary eigenvalues by their analytical prolongation with $\mathcal{E}\rightarrow-i\sqrt{i}\mathcal{E}$ and $T\rightarrow\sqrt{i} T$. For finite time $T$ one can find the RHO by calculating the convolution integral  
\begin{equation}
\begin{aligned}
\mathcal{R}(t)&=\int_{-\infty}^{+\infty}R(t-t')|\psi(t')|^2dt'=\\
&=\frac{e^{\frac{1-4 t T}{4T^2}}}{2T}\left[1+\mbox{Erf}\left(t-\frac{1}{2T}\right) \right].
\label{conv}
\end{aligned}
\end{equation}
This function has a maximum at $t=\bar t(T)$ as shown in the inset in Fig. \ref{fig:1}(c). Hence, for $t\simeq \bar t$ one can write 
$R(t)\simeq R(\bar t)+\frac{1}{2} R^{(2)}(\bar t)(t-\bar t)^2$. The model can be approximated by an RHO as in the spatial case. The temporal decay coefficient is $\gamma_t =|R^{(2)}(\bar t)|$. As a result, one has squeezing also in the temporal case as shown in the quadratures in Fig. \ref{fig:1}(c) that strongly resemble the spatial dynamics in Fig. \ref{fig:1}(b). 

In order to further verify the squeezing, we study the evolution of the phase-space distribution for both the spatial and temporal cases. For squeezed light, the phase-space distribution is elliptical, i.e., it is compressed in the direction of the squeezed variable. \cite{Balazs1990,Barton1986}
The Wigner function $W(x,k)$ 
\begin{equation}
W(x,k)=\frac{1}{\pi}\int \psi^*(x+y)\psi(x-y)e^{2iky} dy
\label{wigner}
\end{equation}
furnishes the phase-space distribution, where $x$ is either the spatial or temporal coordinate, while $k$ is the correspondent conjugated canonical variable. Panels (a) and (c) of Fig. \ref{fig:2} report the simulation of the beam evolution for input power $P=100$ after Eqs. (\ref{nls_ad}) and (\ref{tnls}) respectively. Panels (b) and (d) show the correspondent Wigner distribution $W(x,k)$ at $z=2$ and $z=0$ in the inset. After few steps of propagation $W(x,k)$ becomes elongated and compressed at a slanted direction. We point out that the spectral content is centered around $k\neq 0$.
The elongation results symmetric in the spatial case, while it is asymmetrical in the temporal domain. This asymmetry is due to the causality in the time-response function. The degree of squeezing reaches a maximum value and stops increasing when the HNA does not hold true anymore.\\
These findings can be experimentally tested, by letting a Gaussian laser beam with fixed waist propagate in a nonlinear nonlocal medium, as for example a thermal medium \cite{Gentilini2015glauber,Gentilini2015}. Arranging the setup in order to measure the quadratures $\hat u$ and $\hat v$, a specific value of the beam power exists at which the degree of squeezing is maximum. This happens because the squeezed quadratures rotates in the phase-space with the beam power. A recent paper\cite{Schmidt2017} demonstrate highly non-instantaneous Raman response and the excitation related temporal solitons in liquid-filled hollow-core fibers. These results addresses the effective possibility of exciting squeezing in nonlinear nonlocal media.

In conclusion, squeezing emerges during the propagation of a photon fluid both in temporally and spatially nonlocal media. 
The spectral theory of the squeeze ope\-rator based on Gamow eigenvectors of a reversed harmonic oscillator in a rigged Hilbert space explains the process.
During the evolution, a maximal squeezing is reached until the highly nonlinear approximation is valid. Numerical simulations and the study of the Wigner transform confirm the theory.
Despite the analysis is limited to the classical regime, we have evidence that nonlinear propagation fosters the generation of highly non-Gaussian states that might be employed
for novel quantum-inspired technologies. The implications at a fully quantum level are unknown, and will be deepened in future work. Our work establishes a link between the classical photon fluid description and quantum optics that may potentially surpass limits of squeezing generation for various application as quantum information and gravitational waves detection.

This publication was made possible through the support of a grant from the John Templeton Foundation (58277). The opinions expressed in this publication are those of the authors and do not necessarily reflect the view of the John Templeton Foundation.

%%%%%%%%%%%%%%%%%%%%%%% References %%%%%%%%%%%%%%%%%%%%%%%%%
\bibliography{GIGAbib_agg}% Produces the bibliography via BibTeX.

\begin{thebibliography}{42}%
\makeatletter
\providecommand \@ifxundefined [1]{%
 \@ifx{#1\undefined}
}%
\providecommand \@ifnum [1]{%
 \ifnum #1\expandafter \@firstoftwo
 \else \expandafter \@secondoftwo
 \fi
}%
\providecommand \@ifx [1]{%
 \ifx #1\expandafter \@firstoftwo
 \else \expandafter \@secondoftwo
 \fi
}%
\providecommand \natexlab [1]{#1}%
\providecommand \enquote  [1]{``#1''}%
\providecommand \bibnamefont  [1]{#1}%
\providecommand \bibfnamefont [1]{#1}%
\providecommand \citenamefont [1]{#1}%
\providecommand \href@noop [0]{\@secondoftwo}%
\providecommand \href [0]{\begingroup \@sanitize@url \@href}%
\providecommand \@href[1]{\@@startlink{#1}\@@href}%
\providecommand \@@href[1]{\endgroup#1\@@endlink}%
\providecommand \@sanitize@url [0]{\catcode `\\12\catcode `\$12\catcode
  `\&12\catcode `\#12\catcode `\^12\catcode `\_12\catcode `\%12\relax}%
\providecommand \@@startlink[1]{}%
\providecommand \@@endlink[0]{}%
\providecommand \url  [0]{\begingroup\@sanitize@url \@url }%
\providecommand \@url [1]{\endgroup\@href {#1}{\urlprefix }}%
\providecommand \urlprefix  [0]{URL }%
\providecommand \Eprint [0]{\href }%
\providecommand \doibase [0]{http://dx.doi.org/}%
\providecommand \selectlanguage [0]{\@gobble}%
\providecommand \bibinfo  [0]{\@secondoftwo}%
\providecommand \bibfield  [0]{\@secondoftwo}%
\providecommand \translation [1]{[#1]}%
\providecommand \BibitemOpen [0]{}%
\providecommand \bibitemStop [0]{}%
\providecommand \bibitemNoStop [0]{.\EOS\space}%
\providecommand \EOS [0]{\spacefactor3000\relax}%
\providecommand \BibitemShut  [1]{\csname bibitem#1\endcsname}%
\let\auto@bib@innerbib\@empty
%</preamble>
\bibitem [{\citenamefont {Carusotto}\ and\ \citenamefont
  {Ciuti}(2013)}]{Carusotto2013}%
  \BibitemOpen
  \bibfield  {author} {\bibinfo {author} {\bibfnamefont {I.}~\bibnamefont
  {Carusotto}}\ and\ \bibinfo {author} {\bibfnamefont {C.}~\bibnamefont
  {Ciuti}},\ }\href {\doibase 10.1103/RevModPhys.85.299} {\bibfield  {journal}
  {\bibinfo  {journal} {Rev. Mod. Phys.}\ }\textbf {\bibinfo {volume} {85}},\
  \bibinfo {pages} {299–366} (\bibinfo {year} {2013})}\BibitemShut {NoStop}%
\bibitem [{\citenamefont {Vocke}\ \emph {et~al.}(2015)\citenamefont {Vocke},
  \citenamefont {Roger}, \citenamefont {Marino}, \citenamefont {Wright},
  \citenamefont {Carusotto}, \citenamefont {Clerici},\ and\ \citenamefont
  {Faccio}}]{Vocke15}%
  \BibitemOpen
  \bibfield  {author} {\bibinfo {author} {\bibfnamefont {D.}~\bibnamefont
  {Vocke}}, \bibinfo {author} {\bibfnamefont {T.}~\bibnamefont {Roger}},
  \bibinfo {author} {\bibfnamefont {F.}~\bibnamefont {Marino}}, \bibinfo
  {author} {\bibfnamefont {E.~M.}\ \bibnamefont {Wright}}, \bibinfo {author}
  {\bibfnamefont {I.}~\bibnamefont {Carusotto}}, \bibinfo {author}
  {\bibfnamefont {M.}~\bibnamefont {Clerici}}, \ and\ \bibinfo {author}
  {\bibfnamefont {D.}~\bibnamefont {Faccio}},\ }\href {\doibase
  10.1364/OPTICA.2.000484} {\bibfield  {journal} {\bibinfo  {journal} {Optica}\
  }\textbf {\bibinfo {volume} {2}},\ \bibinfo {pages} {484} (\bibinfo {year}
  {2015})}\BibitemShut {NoStop}%
\bibitem [{\citenamefont {Carusotto}(2014)}]{Carusotto2014}%
  \BibitemOpen
  \bibfield  {author} {\bibinfo {author} {\bibfnamefont {I.}~\bibnamefont
  {Carusotto}},\ }\href@noop {} {\bibfield  {journal} {\bibinfo  {journal}
  {Proc. R. Soc. A}\ }\textbf {\bibinfo {volume} {470}},\ \bibinfo {pages}
  {20140320} (\bibinfo {year} {2014})}\BibitemShut {NoStop}%
\bibitem [{\citenamefont {Larr\'e}\ and\ \citenamefont
  {Carusotto}(2015)}]{Carusotto2015}%
  \BibitemOpen
  \bibfield  {author} {\bibinfo {author} {\bibfnamefont {P.-E.}\ \bibnamefont
  {Larr\'e}}\ and\ \bibinfo {author} {\bibfnamefont {I.}~\bibnamefont
  {Carusotto}},\ }\href {\doibase 10.1103/PhysRevA.92.043802} {\bibfield
  {journal} {\bibinfo  {journal} {Phys. Rev. A}\ }\textbf {\bibinfo {volume}
  {92}},\ \bibinfo {pages} {043802} (\bibinfo {year} {2015})}\BibitemShut
  {NoStop}%
\bibitem [{\citenamefont {Klaers}\ \emph
  {et~al.}(2010{\natexlab{a}})\citenamefont {Klaers}, \citenamefont {Schmitt},
  \citenamefont {Vewinger},\ and\ \citenamefont {Weitz}}]{Weitz2010}%
  \BibitemOpen
  \bibfield  {author} {\bibinfo {author} {\bibfnamefont {J.}~\bibnamefont
  {Klaers}}, \bibinfo {author} {\bibfnamefont {J.}~\bibnamefont {Schmitt}},
  \bibinfo {author} {\bibfnamefont {F.}~\bibnamefont {Vewinger}}, \ and\
  \bibinfo {author} {\bibfnamefont {M.}~\bibnamefont {Weitz}},\ }\href@noop {}
  {\bibfield  {journal} {\bibinfo  {journal} {Nature}\ }\textbf {\bibinfo
  {volume} {468}},\ \bibinfo {pages} {545–548} (\bibinfo {year}
  {2010}{\natexlab{a}})}\BibitemShut {NoStop}%
\bibitem [{\citenamefont {Klaers}\ \emph {et~al.}(2012)\citenamefont {Klaers},
  \citenamefont {Schmitt}, \citenamefont {Damm}, \citenamefont {Vewinger},\
  and\ \citenamefont {Weitz}}]{Weitz2012}%
  \BibitemOpen
  \bibfield  {author} {\bibinfo {author} {\bibfnamefont {J.}~\bibnamefont
  {Klaers}}, \bibinfo {author} {\bibfnamefont {J.}~\bibnamefont {Schmitt}},
  \bibinfo {author} {\bibfnamefont {T.}~\bibnamefont {Damm}}, \bibinfo {author}
  {\bibfnamefont {F.}~\bibnamefont {Vewinger}}, \ and\ \bibinfo {author}
  {\bibfnamefont {M.}~\bibnamefont {Weitz}},\ }\href {\doibase
  10.1103/PhysRevLett.108.160403} {\bibfield  {journal} {\bibinfo  {journal}
  {Phys. Rev. Lett.}\ }\textbf {\bibinfo {volume} {108}},\ \bibinfo {pages}
  {160403} (\bibinfo {year} {2012})}\BibitemShut {NoStop}%
\bibitem [{\citenamefont {Calvanese~Strinati}\ and\ \citenamefont
  {Conti}(2014)}]{Calvanese2014}%
  \BibitemOpen
  \bibfield  {author} {\bibinfo {author} {\bibfnamefont {M.}~\bibnamefont
  {Calvanese~Strinati}}\ and\ \bibinfo {author} {\bibfnamefont
  {C.}~\bibnamefont {Conti}},\ }\href {\doibase 10.1103/PhysRevA.90.043853}
  {\bibfield  {journal} {\bibinfo  {journal} {Phys. Rev. A}\ }\textbf {\bibinfo
  {volume} {90}},\ \bibinfo {pages} {043853} (\bibinfo {year}
  {2014})}\BibitemShut {NoStop}%
\bibitem [{\citenamefont {Conti}\ \emph
  {et~al.}(2010{\natexlab{a}})\citenamefont {Conti}, \citenamefont {Stark},
  \citenamefont {Russell},\ and\ \citenamefont
  {Biancalana}}]{ContiBiancalana2010}%
  \BibitemOpen
  \bibfield  {author} {\bibinfo {author} {\bibfnamefont {C.}~\bibnamefont
  {Conti}}, \bibinfo {author} {\bibfnamefont {S.}~\bibnamefont {Stark}},
  \bibinfo {author} {\bibfnamefont {P.~S.~J.}\ \bibnamefont {Russell}}, \ and\
  \bibinfo {author} {\bibfnamefont {F.}~\bibnamefont {Biancalana}},\ }\href
  {\doibase 10.1103/PhysRevA.82.013838} {\bibfield  {journal} {\bibinfo
  {journal} {Phys. Rev. A}\ }\textbf {\bibinfo {volume} {82}},\ \bibinfo
  {pages} {013838} (\bibinfo {year} {2010}{\natexlab{a}})}\BibitemShut
  {NoStop}%
\bibitem [{\citenamefont {Ghofraniha}\ \emph {et~al.}(2007)\citenamefont
  {Ghofraniha}, \citenamefont {Conti}, \citenamefont {Ruocco},\ and\
  \citenamefont {Trillo}}]{Ghofraniha07}%
  \BibitemOpen
  \bibfield  {author} {\bibinfo {author} {\bibfnamefont {N.}~\bibnamefont
  {Ghofraniha}}, \bibinfo {author} {\bibfnamefont {C.}~\bibnamefont {Conti}},
  \bibinfo {author} {\bibfnamefont {G.}~\bibnamefont {Ruocco}}, \ and\ \bibinfo
  {author} {\bibfnamefont {S.}~\bibnamefont {Trillo}},\ }\href {\doibase
  10.1103/PhysRevLett.99.043903} {\bibfield  {journal} {\bibinfo  {journal}
  {Phys. Rev. Lett.}\ }\textbf {\bibinfo {volume} {99}},\ \bibinfo {eid}
  {043903} (\bibinfo {year} {2007})}\BibitemShut {NoStop}%
\bibitem [{\citenamefont {Mecozzi}\ \emph {et~al.}(1996)\citenamefont
  {Mecozzi}, \citenamefont {Midrio},\ and\ \citenamefont
  {Romagnoli}}]{Mecozzi:96}%
  \BibitemOpen
  \bibfield  {author} {\bibinfo {author} {\bibfnamefont {A.}~\bibnamefont
  {Mecozzi}}, \bibinfo {author} {\bibfnamefont {M.}~\bibnamefont {Midrio}}, \
  and\ \bibinfo {author} {\bibfnamefont {M.}~\bibnamefont {Romagnoli}},\ }\href
  {\doibase 10.1364/OL.21.000402} {\bibfield  {journal} {\bibinfo  {journal}
  {Opt. Lett.}\ }\textbf {\bibinfo {volume} {21}},\ \bibinfo {pages} {402}
  (\bibinfo {year} {1996})}\BibitemShut {NoStop}%
\bibitem [{\citenamefont {Liebling}\ and\ \citenamefont
  {Palenzuela}(2012)}]{Liebling}%
  \BibitemOpen
  \bibfield  {author} {\bibinfo {author} {\bibfnamefont {S.}~\bibnamefont
  {Liebling}}\ and\ \bibinfo {author} {\bibfnamefont {C.}~\bibnamefont
  {Palenzuela}},\ }\href {\doibase 10.1103/PhysRevB.62.1516} {\bibfield
  {journal} {\bibinfo  {journal} {Living Rev. Relativ.}\ }\textbf {\bibinfo
  {volume} {15}} (\bibinfo {year} {2012}),\
  10.1103/PhysRevB.62.1516}\BibitemShut {NoStop}%
\bibitem [{\citenamefont {Roger}\ \emph {et~al.}(2016)\citenamefont {Roger},
  \citenamefont {Maitland}, \citenamefont {Wilson}, \citenamefont {Westerberg},
  \citenamefont {Vocke}, \citenamefont {Wright},\ and\ \citenamefont
  {Faccio}}]{Faccio2016natcom}%
  \BibitemOpen
  \bibfield  {author} {\bibinfo {author} {\bibfnamefont {T.}~\bibnamefont
  {Roger}}, \bibinfo {author} {\bibfnamefont {C.}~\bibnamefont {Maitland}},
  \bibinfo {author} {\bibfnamefont {K.}~\bibnamefont {Wilson}}, \bibinfo
  {author} {\bibfnamefont {N.}~\bibnamefont {Westerberg}}, \bibinfo {author}
  {\bibfnamefont {D.}~\bibnamefont {Vocke}}, \bibinfo {author} {\bibfnamefont
  {E.}~\bibnamefont {Wright}}, \ and\ \bibinfo {author} {\bibfnamefont
  {D.}~\bibnamefont {Faccio}},\ }\href@noop {} {\bibfield  {journal} {\bibinfo
  {journal} {Nature Communications}\ }\textbf {\bibinfo {volume} {7}} (\bibinfo
  {year} {2016})}\BibitemShut {NoStop}%
\bibitem [{\citenamefont {Sun}\ \emph {et~al.}(2012)\citenamefont {Sun},
  \citenamefont {Jia}, \citenamefont {Barsi}, \citenamefont {Rica},
  \citenamefont {Picozzi},\ and\ \citenamefont {Fleischer}}]{Sun2012}%
  \BibitemOpen
  \bibfield  {author} {\bibinfo {author} {\bibfnamefont {C.}~\bibnamefont
  {Sun}}, \bibinfo {author} {\bibfnamefont {S.}~\bibnamefont {Jia}}, \bibinfo
  {author} {\bibfnamefont {C.}~\bibnamefont {Barsi}}, \bibinfo {author}
  {\bibfnamefont {S.}~\bibnamefont {Rica}}, \bibinfo {author} {\bibfnamefont
  {A.}~\bibnamefont {Picozzi}}, \ and\ \bibinfo {author} {\bibfnamefont
  {J.~W.}\ \bibnamefont {Fleischer}},\ }\href
  {http://dx.doi.org/10.1038/nphys2278} {\bibfield  {journal} {\bibinfo
  {journal} {Nat. Phys.}\ }\textbf {\bibinfo {volume} {8}},\ \bibinfo {pages}
  {471–475} (\bibinfo {year} {2012})}\BibitemShut {NoStop}%
\bibitem [{\citenamefont {Picozzi}\ \emph {et~al.}(2006)\citenamefont
  {Picozzi}, \citenamefont {Haelterman}, \citenamefont {Pitois},\ and\
  \citenamefont {Millot}}]{Picozzi06}%
  \BibitemOpen
  \bibfield  {author} {\bibinfo {author} {\bibfnamefont {A.}~\bibnamefont
  {Picozzi}}, \bibinfo {author} {\bibfnamefont {M.}~\bibnamefont {Haelterman}},
  \bibinfo {author} {\bibfnamefont {S.}~\bibnamefont {Pitois}}, \ and\ \bibinfo
  {author} {\bibfnamefont {G.}~\bibnamefont {Millot}},\ }\href@noop {}
  {\bibfield  {journal} {\bibinfo  {journal} {Journal de Physique IV Colloque}\
  }\textbf {\bibinfo {volume} {135}},\ \bibinfo {pages} {33} (\bibinfo {year}
  {2006})}\BibitemShut {NoStop}%
\bibitem [{\citenamefont {Klaers}\ \emph
  {et~al.}(2010{\natexlab{b}})\citenamefont {Klaers}, \citenamefont {Schmitt},
  \citenamefont {Vewinger},\ and\ \citenamefont {Weitz}}]{WeitzNAT2010}%
  \BibitemOpen
  \bibfield  {author} {\bibinfo {author} {\bibfnamefont {J.}~\bibnamefont
  {Klaers}}, \bibinfo {author} {\bibfnamefont {J.}~\bibnamefont {Schmitt}},
  \bibinfo {author} {\bibfnamefont {F.}~\bibnamefont {Vewinger}}, \ and\
  \bibinfo {author} {\bibfnamefont {M.}~\bibnamefont {Weitz}},\ }\href@noop {}
  {\bibfield  {journal} {\bibinfo  {journal} {Nature}\ }\textbf {\bibinfo
  {volume} {468}},\ \bibinfo {pages} {545} (\bibinfo {year}
  {2010}{\natexlab{b}})}\BibitemShut {NoStop}%
\bibitem [{\citenamefont {Klaers}\ \emph
  {et~al.}(2010{\natexlab{c}})\citenamefont {Klaers}, \citenamefont
  {Vewinger},\ and\ \citenamefont {Weitz}}]{WeitzNatPhys2010}%
  \BibitemOpen
  \bibfield  {author} {\bibinfo {author} {\bibfnamefont {J.}~\bibnamefont
  {Klaers}}, \bibinfo {author} {\bibfnamefont {F.}~\bibnamefont {Vewinger}}, \
  and\ \bibinfo {author} {\bibfnamefont {M.}~\bibnamefont {Weitz}},\
  }\href@noop {} {\bibfield  {journal} {\bibinfo  {journal} {Nat. Phys.}\
  }\textbf {\bibinfo {volume} {6}},\ \bibinfo {pages} {512} (\bibinfo {year}
  {2010}{\natexlab{c}})}\BibitemShut {NoStop}%
\bibitem [{\citenamefont {Connaughton}\ \emph {et~al.}(2005)\citenamefont
  {Connaughton}, \citenamefont {Josserand}, \citenamefont {Picozzi},
  \citenamefont {Pomeau},\ and\ \citenamefont {Rica}}]{connaughton:263901}%
  \BibitemOpen
  \bibfield  {author} {\bibinfo {author} {\bibfnamefont {C.}~\bibnamefont
  {Connaughton}}, \bibinfo {author} {\bibfnamefont {C.}~\bibnamefont
  {Josserand}}, \bibinfo {author} {\bibfnamefont {A.}~\bibnamefont {Picozzi}},
  \bibinfo {author} {\bibfnamefont {Y.}~\bibnamefont {Pomeau}}, \ and\ \bibinfo
  {author} {\bibfnamefont {S.}~\bibnamefont {Rica}},\ }\href@noop {} {\bibfield
   {journal} {\bibinfo  {journal} {Phys. Rev. Lett.}\ }\textbf {\bibinfo
  {volume} {95}},\ \bibinfo {pages} {263901} (\bibinfo {year}
  {2005})}\BibitemShut {NoStop}%
\bibitem [{\citenamefont {Walls}(1983)}]{Walls1983}%
  \BibitemOpen
  \bibfield  {author} {\bibinfo {author} {\bibfnamefont {D.}~\bibnamefont
  {Walls}},\ }\href@noop {} {\bibfield  {journal} {\bibinfo  {journal}
  {Nature}\ }\textbf {\bibinfo {volume} {141}} (\bibinfo {year}
  {1983})}\BibitemShut {NoStop}%
\bibitem [{\citenamefont {Walls}\ and\ \citenamefont
  {Milburn}(1999)}]{WallsBook}%
  \BibitemOpen
  \bibfield  {author} {\bibinfo {author} {\bibfnamefont {D.}~\bibnamefont
  {Walls}}\ and\ \bibinfo {author} {\bibfnamefont {G.}~\bibnamefont
  {Milburn}},\ }\href@noop {} {\emph {\bibinfo {title} {{Quantum Optics}}}}\
  (\bibinfo  {publisher} {Spinger-Verlag, Berlin},\ \bibinfo {year}
  {1999})\BibitemShut {NoStop}%
\bibitem [{\citenamefont {Kennard}(1927)}]{Kennard27}%
  \BibitemOpen
  \bibfield  {author} {\bibinfo {author} {\bibfnamefont {E.}~\bibnamefont
  {Kennard}},\ }\href@noop {} {\bibfield  {journal} {\bibinfo  {journal} {Z.
  Phys.}\ }\textbf {\bibinfo {volume} {326}} (\bibinfo {year}
  {1927})}\BibitemShut {NoStop}%
\bibitem [{\citenamefont {Stoler}(1970)}]{Stoler1970}%
  \BibitemOpen
  \bibfield  {author} {\bibinfo {author} {\bibfnamefont {D.}~\bibnamefont
  {Stoler}},\ }\href {\doibase 10.1103/PhysRevD.1.3217} {\bibfield  {journal}
  {\bibinfo  {journal} {Phys. Rev. D}\ }\textbf {\bibinfo {volume} {1}},\
  \bibinfo {pages} {3217} (\bibinfo {year} {1970})}\BibitemShut {NoStop}%
\bibitem [{\citenamefont {Stoler}(1971)}]{Stoler1971}%
  \BibitemOpen
  \bibfield  {author} {\bibinfo {author} {\bibfnamefont {D.}~\bibnamefont
  {Stoler}},\ }\href {\doibase 10.1103/PhysRevD.4.1925} {\bibfield  {journal}
  {\bibinfo  {journal} {Phys. Rev. D}\ }\textbf {\bibinfo {volume} {4}},\
  \bibinfo {pages} {1925} (\bibinfo {year} {1971})}\BibitemShut {NoStop}%
\bibitem [{\citenamefont {Fisher}\ \emph {et~al.}(1984)\citenamefont {Fisher},
  \citenamefont {Nieto},\ and\ \citenamefont {Sandberg}}]{Fisher1984}%
  \BibitemOpen
  \bibfield  {author} {\bibinfo {author} {\bibfnamefont {R.~A.}\ \bibnamefont
  {Fisher}}, \bibinfo {author} {\bibfnamefont {M.~M.}\ \bibnamefont {Nieto}}, \
  and\ \bibinfo {author} {\bibfnamefont {V.~D.}\ \bibnamefont {Sandberg}},\
  }\href {\doibase 10.1103/PhysRevD.29.1107} {\bibfield  {journal} {\bibinfo
  {journal} {Phys. Rev. D}\ }\textbf {\bibinfo {volume} {29}},\ \bibinfo
  {pages} {1107} (\bibinfo {year} {1984})}\BibitemShut {NoStop}%
\bibitem [{\citenamefont {Ma}\ and\ \citenamefont {Rhodes}(1990)}]{Ma1990}%
  \BibitemOpen
  \bibfield  {author} {\bibinfo {author} {\bibfnamefont {X.}~\bibnamefont
  {Ma}}\ and\ \bibinfo {author} {\bibfnamefont {W.}~\bibnamefont {Rhodes}},\
  }\href {\doibase 10.1103/PhysRevA.41.4625} {\bibfield  {journal} {\bibinfo
  {journal} {Phys. Rev. A}\ }\textbf {\bibinfo {volume} {41}},\ \bibinfo
  {pages} {4625} (\bibinfo {year} {1990})}\BibitemShut {NoStop}%
\bibitem [{\citenamefont {Gisin}\ \emph {et~al.}(2002)\citenamefont {Gisin},
  \citenamefont {Ribordy}, \citenamefont {Tittel},\ and\ \citenamefont
  {Zbinden}}]{Gisin2002}%
  \BibitemOpen
  \bibfield  {author} {\bibinfo {author} {\bibfnamefont {N.}~\bibnamefont
  {Gisin}}, \bibinfo {author} {\bibfnamefont {G.}~\bibnamefont {Ribordy}},
  \bibinfo {author} {\bibfnamefont {W.}~\bibnamefont {Tittel}}, \ and\ \bibinfo
  {author} {\bibfnamefont {H.}~\bibnamefont {Zbinden}},\ }\href {\doibase
  10.1103/RevModPhys.74.145} {\bibfield  {journal} {\bibinfo  {journal} {Rev.
  Mod. Phys.}\ }\textbf {\bibinfo {volume} {74}},\ \bibinfo {pages} {145}
  (\bibinfo {year} {2002})}\BibitemShut {NoStop}%
\bibitem [{\citenamefont {Ladd}\ \emph {et~al.}(2010)\citenamefont {Ladd},
  \citenamefont {Jelezko}, \citenamefont {Laflamme}, \citenamefont {Nakamura},
  \citenamefont {Monroe},\ and\ \citenamefont {O’Brien10}}]{Ladd2010}%
  \BibitemOpen
  \bibfield  {author} {\bibinfo {author} {\bibfnamefont {T.~D.}\ \bibnamefont
  {Ladd}}, \bibinfo {author} {\bibfnamefont {F.}~\bibnamefont {Jelezko}},
  \bibinfo {author} {\bibfnamefont {R.}~\bibnamefont {Laflamme}}, \bibinfo
  {author} {\bibfnamefont {Y.}~\bibnamefont {Nakamura}}, \bibinfo {author}
  {\bibfnamefont {C.}~\bibnamefont {Monroe}}, \ and\ \bibinfo {author}
  {\bibfnamefont {J.~L.}\ \bibnamefont {O’Brien10}},\ }\href@noop {}
  {\bibfield  {journal} {\bibinfo  {journal} {Nature}\ }\textbf {\bibinfo
  {volume} {464}},\ \bibinfo {pages} {45–53} (\bibinfo {year}
  {2010})}\BibitemShut {NoStop}%
\bibitem [{\citenamefont {Collaborationa}(2011)}]{Ligo2011}%
  \BibitemOpen
  \bibfield  {author} {\bibinfo {author} {\bibfnamefont {L.~S.}\ \bibnamefont
  {Collaborationa}},\ }\href@noop {} {\bibfield  {journal} {\bibinfo  {journal}
  {Nat. Phys.}\ }\textbf {\bibinfo {volume} {7}},\ \bibinfo {pages} {962–965}
  (\bibinfo {year} {2011})}\BibitemShut {NoStop}%
\bibitem [{\citenamefont {Chua}\ \emph {et~al.}(2014)\citenamefont {Chua},
  \citenamefont {Slagmolen}, \citenamefont {Shaddock},\ and\ \citenamefont
  {McClelland}}]{Chua2014}%
  \BibitemOpen
  \bibfield  {author} {\bibinfo {author} {\bibfnamefont {S.~S.~Y.}\
  \bibnamefont {Chua}}, \bibinfo {author} {\bibfnamefont {B.~J.~J.}\
  \bibnamefont {Slagmolen}}, \bibinfo {author} {\bibfnamefont {D.~A.}\
  \bibnamefont {Shaddock}}, \ and\ \bibinfo {author} {\bibfnamefont {D.~E.}\
  \bibnamefont {McClelland}},\ }\href
  {http://stacks.iop.org/0264-9381/31/i=18/a=183001} {\bibfield  {journal}
  {\bibinfo  {journal} {Classical and Quantum Gravity}\ }\textbf {\bibinfo
  {volume} {31}},\ \bibinfo {pages} {183001} (\bibinfo {year}
  {2014})}\BibitemShut {NoStop}%
\bibitem [{\citenamefont {Schnabel}\ \emph {et~al.}(2010)\citenamefont
  {Schnabel}, \citenamefont {Mavalvala}, \citenamefont {E.},\ and\
  \citenamefont {Lam}}]{Shnabel2010}%
  \BibitemOpen
  \bibfield  {author} {\bibinfo {author} {\bibfnamefont {R.}~\bibnamefont
  {Schnabel}}, \bibinfo {author} {\bibfnamefont {N.}~\bibnamefont {Mavalvala}},
  \bibinfo {author} {\bibfnamefont {M.~D.}\ \bibnamefont {E.}}, \ and\ \bibinfo
  {author} {\bibfnamefont {P.}~\bibnamefont {Lam}},\ }\href@noop {} {\bibfield
  {journal} {\bibinfo  {journal} {Nat. Comm.}\ }\textbf {\bibinfo {volume} {1}}
  (\bibinfo {year} {2010})}\BibitemShut {NoStop}%
\bibitem [{\citenamefont {Vahlbruch}\ \emph {et~al.}(2016)\citenamefont
  {Vahlbruch}, \citenamefont {Mehmet}, \citenamefont {Danzmann},\ and\
  \citenamefont {Schnabel}}]{Vahlbruch2016}%
  \BibitemOpen
  \bibfield  {author} {\bibinfo {author} {\bibfnamefont {H.}~\bibnamefont
  {Vahlbruch}}, \bibinfo {author} {\bibfnamefont {M.}~\bibnamefont {Mehmet}},
  \bibinfo {author} {\bibfnamefont {K.}~\bibnamefont {Danzmann}}, \ and\
  \bibinfo {author} {\bibfnamefont {R.}~\bibnamefont {Schnabel}},\ }\href
  {\doibase 10.1103/PhysRevLett.117.110801} {\bibfield  {journal} {\bibinfo
  {journal} {Phys. Rev. Lett.}\ }\textbf {\bibinfo {volume} {117}},\ \bibinfo
  {pages} {110801} (\bibinfo {year} {2016})}\BibitemShut {NoStop}%
\bibitem [{\citenamefont {Gamow}(1928)}]{gamow1928ZurQuadesAto}%
  \BibitemOpen
  \bibfield  {author} {\bibinfo {author} {\bibfnamefont {G.}~\bibnamefont
  {Gamow}},\ }\href {\doibase 10.1007/BF01343196} {\bibfield  {journal}
  {\bibinfo  {journal} {{Zeitschrift für Physik}}\ }\textbf {\bibinfo {volume}
  {51}},\ \bibinfo {pages} {204} (\bibinfo {year} {1928})}\BibitemShut
  {NoStop}%
\bibitem [{\citenamefont {Chruscinski}(2004)}]{Chruscinski2004squeeze}%
  \BibitemOpen
  \bibfield  {author} {\bibinfo {author} {\bibfnamefont {D.}~\bibnamefont
  {Chruscinski}},\ }\href {\doibase
  http://dx.doi.org/10.1016/j.physleta.2004.05.046} {\bibfield  {journal}
  {\bibinfo  {journal} {Physics Letters A}\ }\textbf {\bibinfo {volume}
  {327}},\ \bibinfo {pages} {290 } (\bibinfo {year} {2004})}\BibitemShut
  {NoStop}%
\bibitem [{\citenamefont {Prigogine}\ \emph {et~al.}(1977)\citenamefont
  {Prigogine}, \citenamefont {Mayné}, \citenamefont {George},\ and\
  \citenamefont {Haan}}]{Prigogine77}%
  \BibitemOpen
  \bibfield  {author} {\bibinfo {author} {\bibfnamefont {I.}~\bibnamefont
  {Prigogine}}, \bibinfo {author} {\bibfnamefont {F.}~\bibnamefont {Mayné}},
  \bibinfo {author} {\bibfnamefont {C.}~\bibnamefont {George}}, \ and\ \bibinfo
  {author} {\bibfnamefont {M.~D.}\ \bibnamefont {Haan}},\ }\href@noop {}
  {\bibfield  {journal} {\bibinfo  {journal} {{Proc. Natl. Acad. Sci. U.S.A.}}\
  }\textbf {\bibinfo {volume} {74}},\ \bibinfo {pages} {4152–6} (\bibinfo
  {year} {1977})}\BibitemShut {NoStop}%
\bibitem [{\citenamefont {Bohm}\ \emph {et~al.}(1999)\citenamefont {Bohm},
  \citenamefont {Scurek},\ and\ \citenamefont
  {Wikramasekara}}]{nucl-th/9902076v1}%
  \BibitemOpen
  \bibfield  {author} {\bibinfo {author} {\bibfnamefont {A.~R.}\ \bibnamefont
  {Bohm}}, \bibinfo {author} {\bibfnamefont {R.}~\bibnamefont {Scurek}}, \ and\
  \bibinfo {author} {\bibfnamefont {S.}~\bibnamefont {Wikramasekara}},\ }\href
  {http://arxiv.org/abs/nucl-th/9902076v1;
  http://arxiv.org/pdf/nucl-th/9902076v1} {\enquote {\bibinfo {title}
  {{Resonances, Gamow Vectors and Time Asymmetric Quantum Theory}},}\ }
  (\bibinfo {year} {1999}),\ \Eprint {http://arxiv.org/abs/nucl-th/9902076v1}
  {arXiv:nucl-th/9902076v1 [nucl-th]} \BibitemShut {NoStop}%
\bibitem [{\citenamefont {Ford}\ and\ \citenamefont
  {Wheeler}(1959)}]{Ford1959}%
  \BibitemOpen
  \bibfield  {author} {\bibinfo {author} {\bibfnamefont {K.~W.}\ \bibnamefont
  {Ford}}\ and\ \bibinfo {author} {\bibfnamefont {J.~A.}\ \bibnamefont
  {Wheeler}},\ }\href@noop {} {\bibfield  {journal} {\bibinfo  {journal} {Ann.
  Phys.}\ }\textbf {\bibinfo {volume} {7}},\ \bibinfo {pages} {259} (\bibinfo
  {year} {1959})}\BibitemShut {NoStop}%
\bibitem [{\citenamefont {Gentilini}\ \emph
  {et~al.}(2015{\natexlab{a}})\citenamefont {Gentilini}, \citenamefont
  {Braidotti}, \citenamefont {Marcucci}, \citenamefont {DelRe},\ and\
  \citenamefont {Conti}}]{Gentilini2015glauber}%
  \BibitemOpen
  \bibfield  {author} {\bibinfo {author} {\bibfnamefont {S.}~\bibnamefont
  {Gentilini}}, \bibinfo {author} {\bibfnamefont {M.~C.}\ \bibnamefont
  {Braidotti}}, \bibinfo {author} {\bibfnamefont {G.}~\bibnamefont {Marcucci}},
  \bibinfo {author} {\bibfnamefont {E.}~\bibnamefont {DelRe}}, \ and\ \bibinfo
  {author} {\bibfnamefont {C.}~\bibnamefont {Conti}},\ }\href@noop {}
  {\bibfield  {journal} {\bibinfo  {journal} {Sci. Rep.}\ }\textbf {\bibinfo
  {volume} {5}} (\bibinfo {year} {2015}{\natexlab{a}})}\BibitemShut {NoStop}%
\bibitem [{\citenamefont {Gentilini}\ \emph
  {et~al.}(2015{\natexlab{b}})\citenamefont {Gentilini}, \citenamefont
  {Braidotti}, \citenamefont {Marcucci}, \citenamefont {DelRe},\ and\
  \citenamefont {Conti}}]{Gentilini2015}%
  \BibitemOpen
  \bibfield  {author} {\bibinfo {author} {\bibfnamefont {S.}~\bibnamefont
  {Gentilini}}, \bibinfo {author} {\bibfnamefont {M.~C.}\ \bibnamefont
  {Braidotti}}, \bibinfo {author} {\bibfnamefont {G.}~\bibnamefont {Marcucci}},
  \bibinfo {author} {\bibfnamefont {E.}~\bibnamefont {DelRe}}, \ and\ \bibinfo
  {author} {\bibfnamefont {C.}~\bibnamefont {Conti}},\ }\href {\doibase
  10.1103/PhysRevA.92.023801} {\bibfield  {journal} {\bibinfo  {journal} {Phys.
  Rev. A}\ }\textbf {\bibinfo {volume} {92}},\ \bibinfo {pages} {023801}
  (\bibinfo {year} {2015}{\natexlab{b}})}\BibitemShut {NoStop}%
\bibitem [{\citenamefont {Marcucci}\ and\ \citenamefont
  {Conti}(2016)}]{Marcucci16}%
  \BibitemOpen
  \bibfield  {author} {\bibinfo {author} {\bibfnamefont {G.}~\bibnamefont
  {Marcucci}}\ and\ \bibinfo {author} {\bibfnamefont {C.}~\bibnamefont
  {Conti}},\ }\href {\doibase 10.1103/PhysRevA.94.052136} {\bibfield  {journal}
  {\bibinfo  {journal} {Phys. Rev. A}\ }\textbf {\bibinfo {volume} {94}},\
  \bibinfo {pages} {052136} (\bibinfo {year} {2016})}\BibitemShut {NoStop}%
\bibitem [{\citenamefont {Conti}\ \emph
  {et~al.}(2010{\natexlab{b}})\citenamefont {Conti}, \citenamefont {Schmidt},
  \citenamefont {Russell},\ and\ \citenamefont
  {Biancalana}}]{Conti2010_linearons}%
  \BibitemOpen
  \bibfield  {author} {\bibinfo {author} {\bibfnamefont {C.}~\bibnamefont
  {Conti}}, \bibinfo {author} {\bibfnamefont {M.~A.}\ \bibnamefont {Schmidt}},
  \bibinfo {author} {\bibfnamefont {P.~S.~J.}\ \bibnamefont {Russell}}, \ and\
  \bibinfo {author} {\bibfnamefont {F.}~\bibnamefont {Biancalana}},\ }\href
  {\doibase 10.1103/PhysRevLett.105.263902} {\bibfield  {journal} {\bibinfo
  {journal} {Phys. Rev. Lett.}\ }\textbf {\bibinfo {volume} {105}},\ \bibinfo
  {pages} {263902} (\bibinfo {year} {2010}{\natexlab{b}})}\BibitemShut
  {NoStop}%
\bibitem [{\citenamefont {Reeves}\ \emph {et~al.}(2003)\citenamefont {Reeves},
  \citenamefont {Skryabin}, \citenamefont {Biancalana}, \citenamefont {Knight},
  \citenamefont {Russell}, \citenamefont {Omenetto}, \citenamefont {A.},\ and\
  \citenamefont {J.}}]{Skryabin03}%
  \BibitemOpen
  \bibfield  {author} {\bibinfo {author} {\bibfnamefont {W.~H.}\ \bibnamefont
  {Reeves}}, \bibinfo {author} {\bibfnamefont {D.~V.}\ \bibnamefont
  {Skryabin}}, \bibinfo {author} {\bibfnamefont {F.}~\bibnamefont
  {Biancalana}}, \bibinfo {author} {\bibfnamefont {J.~C.}\ \bibnamefont
  {Knight}}, \bibinfo {author} {\bibfnamefont {P.~S.~J.}\ \bibnamefont
  {Russell}}, \bibinfo {author} {\bibfnamefont {F.~G.}\ \bibnamefont
  {Omenetto}}, \bibinfo {author} {\bibfnamefont {E.}~\bibnamefont {A.}}, \ and\
  \bibinfo {author} {\bibfnamefont {T.~A.}\ \bibnamefont {J.}},\ }\href@noop {}
  {\bibfield  {journal} {\bibinfo  {journal} {Nature}\ }\textbf {\bibinfo
  {volume} {424}},\ \bibinfo {pages} {511} (\bibinfo {year}
  {2003})}\BibitemShut {NoStop}%
\bibitem [{\citenamefont {Balazs}\ and\ \citenamefont
  {Voros}(1990)}]{Balazs1990}%
  \BibitemOpen
  \bibfield  {author} {\bibinfo {author} {\bibfnamefont {N.~L.}\ \bibnamefont
  {Balazs}}\ and\ \bibinfo {author} {\bibfnamefont {A.}~\bibnamefont {Voros}},\
  }\href@noop {} {\bibfield  {journal} {\bibinfo  {journal} {Ann. Phys.}\
  }\textbf {\bibinfo {volume} {199}},\ \bibinfo {pages} {123} (\bibinfo {year}
  {1990})}\BibitemShut {NoStop}%
\bibitem [{\citenamefont {Barton}(1986)}]{Barton1986}%
  \BibitemOpen
  \bibfield  {author} {\bibinfo {author} {\bibfnamefont {G.}~\bibnamefont
  {Barton}},\ }\href@noop {} {\bibfield  {journal} {\bibinfo  {journal} {Ann.
  Phys.}\ }\textbf {\bibinfo {volume} {166}},\ \bibinfo {pages} {322–363}
  (\bibinfo {year} {1986})}\BibitemShut {NoStop}%
	\bibitem [{\citenamefont {Chemnitz}\ \emph {et~al.}(2017)\citenamefont
  {Chemnitz}, \citenamefont {Gebhardt}, \citenamefont {Gaida}, \citenamefont
  {Stutzki}, \citenamefont {Kobelke}, \citenamefont {Limpert}, \citenamefont
  {Tünnermann},\ and\ \citenamefont {Schmidt}}]{Schmidt2017}%
  \BibitemOpen
  \bibfield  {author} {\bibinfo {author} {\bibfnamefont {M.}~\bibnamefont
  {Chemnitz}}, \bibinfo {author} {\bibfnamefont {M.}~\bibnamefont {Gebhardt}},
  \bibinfo {author} {\bibfnamefont {C.}~\bibnamefont {Gaida}}, \bibinfo
  {author} {\bibfnamefont {F.}~\bibnamefont {Stutzki}}, \bibinfo {author}
  {\bibfnamefont {J.}~\bibnamefont {Kobelke}}, \bibinfo {author} {\bibfnamefont
  {J.}~\bibnamefont {Limpert}}, \bibinfo {author} {\bibfnamefont
  {A.}~\bibnamefont {Tünnermann}}, \ and\ \bibinfo {author} {\bibfnamefont
  {M.~A.}\ \bibnamefont {Schmidt}},\ }\href@noop {}
  {\bibfield  {journal} {\bibinfo  {journal} {Nat. Comm.}\ }\textbf {\bibinfo
  {volume} {8}},\ \bibinfo {pages} {42} (\bibinfo {year} {2017})}\BibitemShut
  {NoStop}%
\end{thebibliography}

\end{document}